\documentclass{iau}
\usepackage{graphicx}
\usepackage{natbib}

\title[Uncertainties in NEO deflection demo missions] 
{Dealing with uncertainties in asteroid deflection demonstration missions: NEOT$\omega$IST}

\author[Eggl et al.]   
{Siegfried Eggl$^1$,
 Daniel Hestroffer$^1$,
 Juan L. Cano$^2$,
 Javier Mart\'in \'Avila$^2$,
 Line Drube$^3$,
 Alan W. Harris$^3$,
 Albert Falke$^4$,
 Ulrich Johann$^4$,
 Kilian Engel$^4$,
 Stephen R. Schwartz$^5$
 \and Patrick Michel$^5$}

\affiliation{$^1$IMCCE Observatoire de Paris, UPMC Paris-06, Univ. Lille 1 \\ 77 av. Denfert-Rochereau, 75014 Paris, France
 \\ email: {\tt siegfried.eggl@obspm.fr} \\[\affilskip]
$^2$ DEIMOS Space S.L.U., Ronda	de Poniente 19, 28760 Tres Cantos, Spain  \\[\affilskip]
$^3$ Institute of Planetary Research, German Aerospace Center (DLR), Rutherfordstr. 2, 12489 Berlin, Germany\\[\affilskip]
$^4$ Airbus DS, Claude-Dornier-Str., 88090 Immenstaad, Germany\\[\affilskip]
$^5$ Labratoire Lagrange, Univ. Nice, CNRS, Observatoire de la C\^ote d\'{ }Azur, Boulevard de l\'{}Observatoire, CS 34229, 06304 Nice Cedex 4, France
}

\pubyear{2015}
\volume{318}  
\pagerange{119--126}
\jname{IAUS 318 - Asteroids: New Observations, New Models}
\editors{S. Chesley, A. Morbidelli, R. Jedicke \& D. Farnocchia, eds.}
\begin{document}

\maketitle

\begin{abstract}
Deflection missions to near-Earth asteroids will encounter non-negligible uncertainties in the physical and orbital parameters of the target object.
In order to reliably assess future impact threat mitigation operations 
such uncertainties have to be quantified and incorporated into the mission design.  
The implementation of deflection demonstration missions offers the great opportunity to test our current understanding of 
deflection relevant uncertainties and their consequences, e.g., regarding
kinetic impacts on asteroid surfaces. 
In this contribution, we discuss the role of uncertainties in the NEOT$\omega$IST asteroid deflection demonstration concept,
a low-cost kinetic impactor design elaborated in the framework of the NEOShield project.
The aim of NEOT$\omega$IST is to change the spin state of a known and well characterized near-Earth object, in this case the asteroid (25143) Itokawa.  
Fast events such as the production of the impact crater and ejecta are studied via cube-sat chasers and a flyby vehicle.
Long term changes, for instance, in the asteroid's spin and orbit, can be assessed using ground based observations.
We find that such a mission can indeed provide valuable constraints on mitigation relevant parameters. Furthermore, the here proposed
kinetic impact scenarios can be implemented within the next two decades without threatening Earth's safety.
\end{abstract}

\firstsection 
\section{Introduction}
Near-Earth objects (NEO) collide with our planet on a regular basis.   
Although current NEO population and impact models suggest that such impacts are unlikely to cause 
global catastrophes in the near future \citep{harris-2015}
it is still reasonable to establish impact threat mitigation options. Even for sub-kilometer sized objects, 
the potential loss of human life as well as the damage to local infrastructure and economy can more than justify an asteroid orbit deflection attempt.
While many deflection options have been considered in the past, it has become more and more clear that the accurate assessment of 
an asteroid's physical and orbital parameters as well as their uncertainties is the key
ingredient to a successful impact threat mitigation, no matter which technique is chosen \citep[][]{sugimoto-et-al-2014}.
Current model predictions are deemed to be sufficiently close to reality to allow for a successful mitigation mission planning. Yet, 
only a practical implementation of a deflection demonstration mission can confirm this belief.
Several innovative asteroid deflection demonstration concepts have, therefore, been presented to space agencies \cite[e.g.][]{donquijote-2006,aida-2012,arm-2014}.

\section{NEOT$\omega$IST}
The NEOShield consortium \citep{harris-et-al-2013} has recently explored low cost deflection demo mission options capable of delivering a reasonable amount of 
data for model validation at
minimum expense. Core to the resulting NEOT$\omega$IST concept \citep{neotwist-2015} is the idea 
to use an off-center kinetic impact to change the spin state of an already well known NEO such as (25143) Itokawa. The main advantage
of this approach lies in the fact that sending an expensive second spacecraft on a rendez-vous trajectory to characterize the target and observe the impact 
can be avoided. 
In fact, short term information such as the ejecta distribution produced by the impact can be observed 
by a fly-by vehicle and/or chasers that decouple from the impactor some time before the collision.
Long term changes in the rotation state caused by the kinetic impact can be reconstructed via ground based observations and 
linked with mitigation parameters to a reasonable degree of accuracy, depending, of course, on the remaining uncertainties in the target's physical parameters.
Conservative estimates predict a delivered impact momentum of roughly $5\times 10^3$ kNs resulting in a change in Itokawa's rotational period of approximately 4 minutes ($\approx0.5\%$). This change is large enough to be validated with a reasonably high `signal-to-noise ratio` (SNR $>10$) within the first year after the impact using ground based
observations alone \citep{neotwist-2015}. However, uncertainties in the targets' physical properties and the impact delivery remain non-negligible, even for a well characterized
target such as (25143) Itokawa.
A conscientious evaluation of uncertainties and their consequences on the deflection mission outcome is, thus, vital, as it directly influences our
ability to derive relevant parameters from observations.
%

\section{The role of uncertainties in deflection demonstration missions}
\begin{figure} 
\begin{center}
     {\includegraphics[angle=0,width=0.6\textwidth]{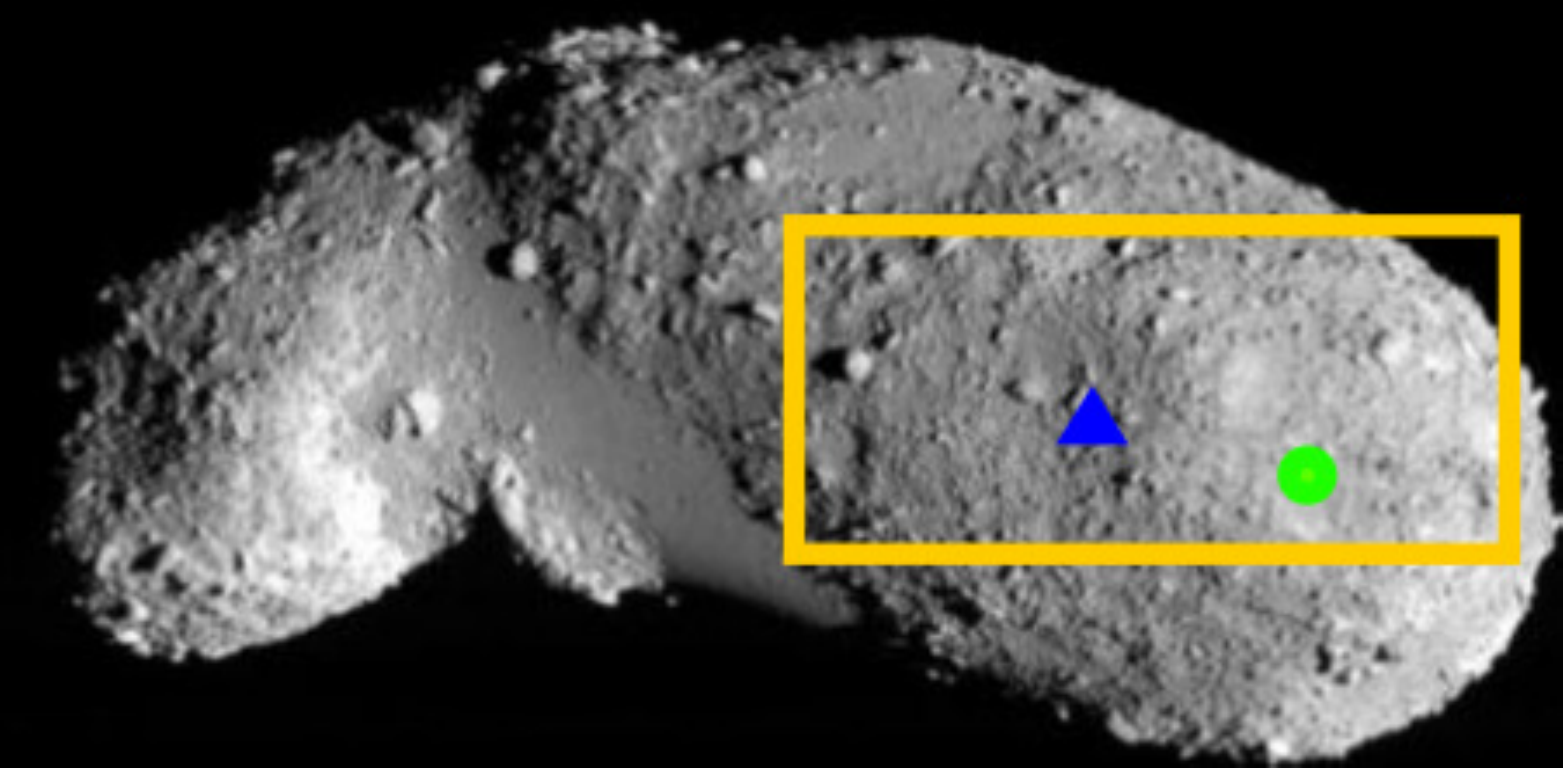}} \\
     {\includegraphics[angle=0,width=1\textwidth]{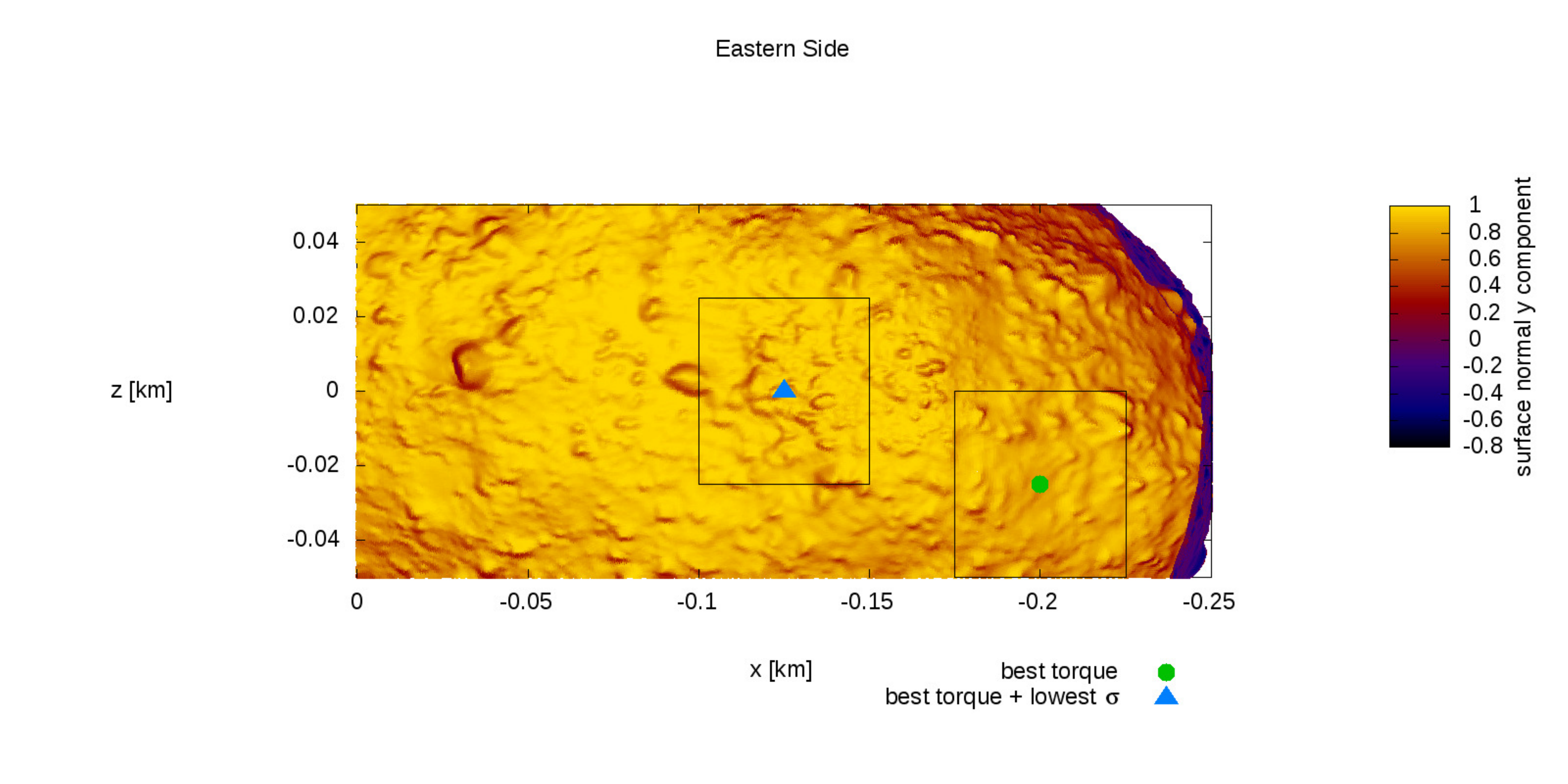}}\\
   \caption{Map of (25143) Itokawa showing probable NEOT$\omega$IST impact locations.
   The preferred impact side (Eastern side) is presented on top.
   A zoom on the area hosting the impact locations (yellow rectangle) is presented
   below.
   The normalized length of the impact-orthogonal components of the surface normal vectors is color coded.
   Positive values point out of the image, negative values into the image.
   The circle and the triangle symbols indicate the best impact locations with respect to optimum
   torque and optimum torque with lowest surface normal variance. 
\label{fig:snorm}}
\end{center}
\end{figure}

Even if it were possible to send explorer spacecraft as part of every deflection mission or deflection mission demonstration, our knowledge of key parameters
such as the asteroid's mass and subsurface structure would never be perfect. This may not be particularly problematic,
as the remaining uncertainties simply need to be accounted for in the deflection planning. 
Depending on the available warning time and resources devoted to an actual deflection attempt, relatively large uncertainties can 
be acceptable as long as deflection success, i.e., a minimum miss distance with respect to the Earth, is guaranteed \citep{sugimoto-et-al-2014}.
How about asteroid deflection demonstration missions? 
For one, uncertainties need to be quantified and, if possible, constrained in order to guarantee a validation of the effect. 
In other words, uncertainties in the target's observables must not be larger than
the change imparted by the demonstration mission. 
If one attempts to derive deflection parameters such as the momentum enhancement factor, $\beta$ \citep{holsapple-housen-2012}, 
from the effective change of a demo mission target's spin state, for instance,
achieving a high SNR between the difference in rotation periods and its uncertainty becomes a necessity. 
A study of uncertainties regarding the impact delivery, Itokawa's topography, and 
the location of its spin axis and consequently its moments of inertia shows that 
a NEOT$\omega$IST concept realization should permit to measure $\beta$ with a relative precision better than $\sigma_{\beta}/\beta \approx 0.3$ 
using ground based observations alone. For more details on this topic we would like to refer the reader to \cite{neotwist-2015}.
Apart from impact validation and parameter estimation, 
uncertainties also play a fundamental role in post mitigation impact risk assessment (PMIRA) \citep{eggl-et-al-2015}. 
Changes in the target's orbit caused by the deflection demonstration are not supposed to increase the impact probability with respect to the Earth 
in the foreseeable future. If a deflection demonstration mission is subject to large uncertainties,  
a risk analysis has to be performed in order to dispel planetary safety concerns. 
Of course, the primary goal of the NEOT$\omega$IST concept is to change the spin state of the target asteroid.
However, an off-center impact will alter the target's heliocentric orbit as well. 
In order to discuss the relevant uncertainties for a kinetic impactor-based mission such as  NEOT$\omega$IST,   
and to see how these may affect the target's heliocentric orbit, let us briefly recall the following equation \citep{scheeres-2015}:
%
%
\begin{equation}
\Delta \vec V\approx \frac{m}{M}[\mathcal I+(\beta-1)\hat n \hat n] \Delta \vec v    \label{eq:dv},
\end{equation}
where $\Delta \vec V$ denotes the change in the asteroid's velocity vector and $\Delta \vec v$ the velocity of the impactor spacecraft relative
to the target. The masses of the asteroid and the spacecraft are represented as $M$ and $m$, respectively. Furthermore, $\hat n$ represents
the asteroid surface normal vector at the impact location (of unit size), $\hat n \hat n$ is the dyadic product of the same and $\mathcal I$ denotes the unit dyade 
(equivalent to the identity matrix, i.e. $\mathcal I\; \vec v = \vec v \; \mathcal I =\vec v$). 
Equation (\ref{eq:dv}) describes the direct momentum transfer due to the kinetic impactor
as well as the indirect momentum transfer due to the ejecta. The latter is assumed to be in the direction of the asteroid surface normal 
at the location of the impact with a magnitude that scales with the momentum enhancement factor $\beta$. 
When $\beta=1$, neither the impactor nor any ejecta escape and only the momentum of the impactor is transferred to the asteroid. 
In case ejecta are produced that are fast enough to escape the asteroid's gravitational well, laboratory experiments and simulations suggest,
that $\beta$ may be substantially larger depending on the impact as well as the regolith, surface and subsurface properties of the target 
\citep{jutzi-michel-2014,holsapple-housen-2012}. 
Some of those material properties were not acquired during the Hayabusa mission \citep{hayabusa-2006} to (25143) Itokawa, the primary NEOT$\omega$IST target, 
and it is very difficult to gather such data from ground-based observations. 
Hence, our estimates of $\beta$ will contain a significant amount of uncertainty. 
In fact, all quantities in equation (\ref{eq:dv}) have uncertainties associated with them. The precision of the 
mass of (25143) Itokawa, for instance, is not better than 3-5\% \citep{fujiwara2006rubble,abe2006mass}.
Uncertainties in the asteroid's velocity, $\vec V$ come form the asteroid's orbit uncertainty at the time of the impact. 
Even the direction of the surface normal, $\hat n$ at the location of the impact is difficult to predict, since limitations in guidance, 
navigation and control (GNC) of the impactor
can only guarantee that the impactor hits the asteroid within a certain area ($25\;$m $\times$ $25\;$m) rather than at a certain spot. As the surface normal orientation varies on
smaller length scales, the direction of the ejecta becomes uncertain; see Figure \ref{fig:snorm}. 
The uncertainties of surface normals in the GNC confidence area have to be taken into account as well to determine, 
for instance, the optimum location for an impact that changes Itokawa's spin. They also add to the control uncertainty associated with a kinetic impact \citep{scheeres-2015}.
The blue triangle and the green circle in Figure \ref{fig:snorm} indicate the difference between impact locations when maximizing only the achievable torque based on the local surface normal
at the impact point (green circle) or when maximizing the achievable torque whilst minimizing the surface normal variance (blue triangle).
In the latter case, a part of the achievable torque is sacrificed to limit uncertainties. For details see \citet{neotwist-2015}. All of the uncertainties that have been described qualitatively in this section
are relevant to assessing both, the probable change in Itokawa's spin state and its heliocentric orbit.
In the following section we aim to investigate the influence of those uncertainties on the future impact risk of (25143) Itokawa.
\section{Post Mitigation Impact Risk Assessment}\label{eggl:sec:pmira}
The goal of this section is to quantify probable uncertainties that occur in five NEOT$\omega$IST mission scenarios
for the target asteroid (25143) Itokawa. Furthermore, we perform an impact risk re-assessment for Itokawa, taking the changes in the asteroid's orbit into account
that have been produced by the kinetic impact.
The five mission scenarios under investigation are summarized in Table \ref{tab:0}. Our post mitigation impact risk assessment (PMIRA)
takes the following uncertainties into account: 
\begin{itemize}
\item Itokawa's mass and orbit uncertainties at the time of the impact,
\item variations in the topography of the impact area,
\item as well as the impactor mass uncertainty that is due to possible malfunctions of the release
mechanisms of the flyby vehicle or cube-sat chasers.
\end{itemize}
The PMIRA process is then performed as follows:  
First, uncertainties relevant to the change of the heliocentric orbit of (25143) Itokawa are quantified.
The resulting uncertainty domain is sampled through many possible realizations (clones) of the dynamical system 
after the kinetic impact. All clones are propagated over a certain time span to check for possible impacts with the Earth.
If no impacts are detected, the statistical changes in the minimum encounter distance (MED) between the asteroid and the Earth are surveyed and discussed.
The relevant parameter ranges and uncertainties can be quantified as follows:
Itokawa's mass range is $3.51\pm0.104 \times 10^{10}$ kg \citep{fujiwara2006rubble}.
The impactor is assumed to have masses between 520 kg and 640 kg (nominal 520 kg) depending on whether or not all observational equipment
that is to be launched prior to the impact is released correctly.
Equation (\ref{eq:dv}) dictates that both, magnitude and direction of the impactor, as well as of the ejecta have to be considered to correctly 
calculate the change in the target asteroid's momentum.
While the direction of the impactor was fixed according to the scenario specifications in Table \ref{tab:0}, 
the direction of the net ejecta momentum vector (EMV) was 
drawn from a uniform distribution in a cone with an opening angle of $2\phi=46$ deg around the local surface normal.
\footnote{Please note that this is not the opening angle
 of the ejecta plume, but the possible range of directions of the net ejecta momentum vector.} 
The latter number was chosen 
in accordance with the variance of the surface normals close to the impactor target areas, see Figure \ref{fig:snorm}. Thus, the uncertainty in the local
topography is accounted for.
The length of the net EMV was scaled to match an overall $\beta$ factor range between 1.2 and 3.5 (nominal 2) \citep{jutzi-michel-2014}.
Orbital elements and their covariances, i.e., uncertainties for (25143) Itokawa, were taken from JPL's small body data base \citep{chamberlin2010jpl}.
The combined covariance matrix of all uncertainties was sampled with a total number of 1000 clones per run along the largest Eigenvector of the covariance matrix
between $\pm3\sigma$ \citep[Extended Line of Variation, e.g.,][]{spoto-et-al-2014}. Orbit uncertainties were assumed to have Gaussian probability densities,
whereas the mitigation relevant parameters were drawn from uniform distributions \citep{eggl-et-al-2015}. 
An overall mission failure probability of 10\% was assumed.
\begin{table}
\begin{center}
\begin{tabular}{l|ccccc}
\hline
 scenario & S1 & S2 & S3 & S4 & S5\\\hline
 year of arrival & 2024 & 2027 & 2030 & 2033  & 2036  \\\hline
swingbys & - & - & - & - & Earth                           \\\hline
flight time (days) & 581 & 942 & 589 & 95 & 503 \\\hline
 Earth $v_{\infty}$ (km/s) &
 2.07 &
 2.39 &
 2.77 &
 2.18 &
 1.62 \\\hline
 Earth $v_\infty$ declination (deg) &
 0.1 &
 0.3 &
 2.7 &
 4.7 &
 -5.0 \\\hline
arrival $v_\infty$ perpendicular to Itokawa axis (km/s) &
 7.99 &
 8.45 &
 8.86 &
 8.97 &
 8.92 \\\hline
arrival $v_\infty$ declin. w.r.t. Itokawa equator (deg) &
 4.5 &
 5.4 &
 5.2 &
 2.6 &
 2.5 \\\hline
payload mass (w/o LPF, kg) &
 424 &
 401 &
 372 &
 416 &
 447 \\\hline
linear impulse normal to Itokawa axis (kN s) &
 5522 &
 5653 &
 5667 &
 6135 &
 6372 \\\hline
 minimum mass linear impulse (kN s) &
 5107 &
 5408 &
 5667 &
 5740 &
 5704 \\\hline
 Sun-Itokawa-impactor angle at impact (deg) &
 30 &
 24 &
 21 &
 12 &
 13 \\\hline
Earth-Itokawa distance at impact (au) & 
 0.95 &
 0.68 &
 0.39 &
 0.09 &
 0.25\\\hline
\end{tabular}
\caption{Proposed NEOT$\omega$IST mission scenarios. The payload mass refers to the achievable arrival mass without the Lisa Path Finder (LPF) upper stage (LPF dry
mass is 268 kg), but the impact linear impulse is computed with the actual arrival mass, which
includes the LPF.
\label{tab:0}}
\end{center}
\end{table}
\begin{table}
\begin{center}
\begin{tabular}{c|c|ccccc}
\hline
Earth encounter & $d_{nom}$  & \multicolumn{5}{c}{$\Delta$ MED [km]}   \\\hline
$[$year$]$ & [au] & S1  & S2  & S3  & S4  & S5  \\\hline
2036 &  0.041 & 10   &  10 & 10  & 1  & -/-   \\ 
2039 &  0.185 & -55  & -51 & -51 & -37  & -28 \\
2068 &  0.228 & -472 & -241 & -40 & 119  & 0 \\\hline
2071 &  0.027 & 70   & 62  &  29 & -72 & -165 \\\hline
2074 &  0.047 & 401  & 401  & 96  & -218 & -505 \\
2100 &  0.128 & 915  & 915 & 208 & -505 &  -1158 \\
2103 &  0.045 & 450  & 450 &  242  &-149  & -522 \\\hline
\end{tabular}
\caption{Changes in the minimum Earth-miss distances due to the kinetic impact scenarios S1-S5. All future encounters between
(25143) Itokawa and the Earth that are closer than 0.25 au up to the year 2105 are considered. $d_{nom}$ denotes the nominal approach distance between Itokawa and the center of gravity of the Earth.
The values in the five scenario columns represent the difference between untouched cases and cases with a kinetic impact. More precisely, the difference between untouched and deflected 
$E(d_{clone}) - 3\sigma$ is shown,
where $E$ is the expected value and $\sigma$ is the square root of the variance of the clone distance $d_{clone}$ to the Earth during the respective close encounter. Negative numbers indicate 
an undesirable shift of the orbit solutions towards the Earth.  
The deepest close encounter (global MED) of Itokawa with the Earth (in 2071) is highlighted.
\label{tab:1}}
\end{center}
\end{table}
\begin{figure} 
\begin{center}
     {\includegraphics[angle=-90,width=0.8\textwidth]{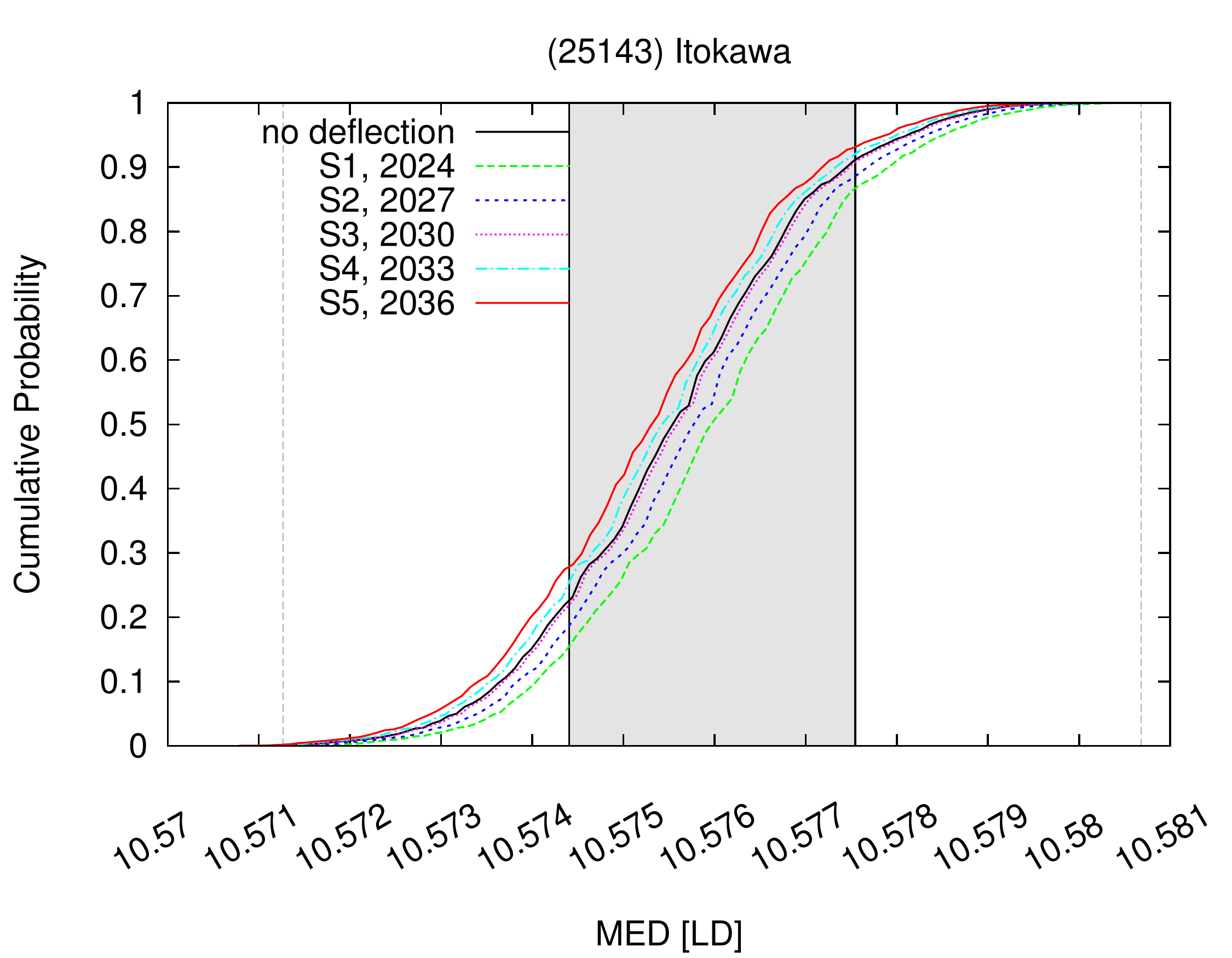}}\\
   \caption{Cumulative probability distributions (CPD) of the closest encounter distance (MED of the year 2071) of (25143) Itokawa with the Earth
    are presented for various NEOT$\omega$IST scenarios. The nominal solution is represented by the black continuous S-shaped line in the center.
   Scenarios 1, 2 and 3 most likely enlarge the (smallest) distance between Itokawa and the Earth in 2071. In contrast, scenarios 4 and 5 bring Itokawa closer to the
   Earth compared to the non-deflected case. The gray area and vertical lines represent the $1\sigma$ and $3\sigma$ non-deflected solutions by JPL, respectively.
   Post kinetic impact solutions with uncertainties are shown. For a description of scenarios 1-5 see Table \ref{tab:0}. 
\label{fig:cpd}}
\end{center}
\end{figure}
First, we considered the non-deflected (untouched) case where only orbit uncertainties play a role. The subsequent calculations contained 
the scenario-derived kinetic impact including all previously described uncertainties. 
All cases were propagated up through January $1^{st}$, 2105. The corresponding dynamical model included the planets of the Solar system,
the Earth-moon, the Pluto-Charon system as well as the BIG16, i.e., the 16 most relevant perturbing asteroids as listed in the JPL-HORIZONS system \citep[e.g.][]{horizons-2015}.
Post-Newtonian corrections were incorporated as discussed in \citet{eggl-et-al-2015}.
Our results are presented in Figure \ref{fig:cpd} as well as in Table \ref{tab:1}.
One can see that none of these scenarios are capable of increasing the Earth-to-Itokawa miss distance for all future close encounters. 
Scenarios 1, 2 and 3 are, nevertheless, acceptable from a planetary safety perspective as they increase the (global) MED between Itokawa and the Earth, i.e., the MED during the closest approach in the year 2071. 
In contrast, the distance between Itokawa and the Earth would most likely shrink, if Scenarios 4 or 5 were chosen. 
The kinetic impact-induced changes in the MED in both cases are of the order of a hundred kilometers
during the close encounter in 2071. As the actual distance between Itokawa and the Earth at that time is more than 10 LD, 
the corresponding Palermo Scale values remain negligibly small.
Scenario 1 would result in the largest increase of the minimum encounter distance between Itokawa and the Earth, while Scenario 5 yields the largest decrease.

\section{Summary and Conclusions}
In this work we have investigated the role of uncertainties in the NEOT$\omega$IST asteroid deflection demonstration concept.
In order to predict changes in the target asteroid's orbit and spin state, uncertainties had to be taken into account. 
Applying a probabilistic approach we have simulated the outcome of five different NEOT$\omega$IST mission scenarios with the aim to see whether the change in the target
asteroid's heliocentric orbit
--- a byproduct of the change of the spin state --- yields an increased risk of collision between (25143) Itokawa and Earth during a future encounter. 
None of the five NEOT$\omega$IST mission designs proposed in this article would threaten planetary safety if implemented.  
Early launch and impact scenarios (S1-S3) even offer the possibility to enlarge 
the close approach distance between Itokawa and Earth slightly in 2071.

\vspace{0.5cm}
\noindent
\textbf{Acknowledgments}\\
This research has received funding from the
European Union's Seventh Framework Program (FP7/2007-2013) under grant
agreement no. 282703 (NEOShield) and H2020-PROTEC-2014 - Protection of European assets in
and from space project no. 640351 (NEOShield-2).


\begin{thebibliography}{17}
\expandafter\ifx\csname natexlab\endcsname\relax\def\natexlab#1{#1}\fi

\bibitem[{Abe {et~al.}(2006)Abe, Mukai, Hirata, Barnouin-Jha, Cheng, Demura,
  Gaskell, Hashimoto, Hiraoka, Honda, {et~al.}}]{abe2006mass}
Abe, S. {et~al.} 2006, \textit{Science}, 312, 1344

\bibitem[{Chamberlin \& Yeomans(2010)}]{chamberlin2010jpl}
Chamberlin, A., \& Yeomans, D. 2010, Jet Propulsion Laboratory: Solar System
  Dynamics. http://ssd. jpl. nasa. gov

\bibitem[{{Cheng} {et~al.}(2012){Cheng}, {Rivkin}, {Galvez}, {Carnelli},
  {Michel}, \& {Reed}}]{aida-2012}
{Cheng}, A.~F., {Rivkin}, A., {Galvez}, A., {Carnelli}, I., {Michel}, P., \&
  {Reed}, C. 2012, in AAS/Division for Planetary Sciences Meeting Abstracts,
  Vol.~44, AAS/Division for Planetary Sciences Meeting Abstracts, 215.03

\bibitem[{{Drube} {et~al.}(2015){Drube}, {Harris}, {Engel}, {Falke}, {Johann},
  {Eggl}, {Cano}, M., R., \& P.}]{neotwist-2015}
{Drube}, L. {et~al.} 2015, \textit{AcA}, submitted

\bibitem[{Eggl {et~al.}(2015)Eggl, Hestroffer, Thuillot, Bancelin, Cano, \&
  Cichocki}]{eggl-et-al-2015}
Eggl, S., Hestroffer, D., Thuillot, W., Bancelin, D., Cano, J.~L., \& Cichocki,
  F. 2015, \textit{Adv. Sp. Res.}, 56, 528 , advances in Asteroid and Space
  Debris Science and Technology - Part 1

\bibitem[{Fujiwara {et~al.}(2006)Fujiwara, Kawaguchi, Yeomans, Abe, Mukai,
  Okada, Saito, Yano, Yoshikawa, Scheeres, {et~al.}}]{fujiwara2006rubble}
Fujiwara, A. {et~al.} 2006, \textit{Science}, 312, 1330

\bibitem[{Giorgini(2015)}]{horizons-2015}
Giorgini, J.~D. 2015, IAU General Assembly, 22, 56293

\bibitem[{Harris {et~al.}(2013)Harris, Barucci, Cano, Fitzsimmons, Fulchignoni,
  Green, Hestroffer, Lappas, Lork, Michel, Morrison, Payson, \&
  Schäfer}]{harris-et-al-2013}
Harris, A. {et~al.} 2013, \textit{AcA}, 90, 80

\bibitem[{{Harris} \& {D'Abramo}(2015)}]{harris-2015}
{Harris}, A.~W., \& {D'Abramo}, G. 2015, Icarus, 257, 302

\bibitem[{{Harris} {et~al.}(2006){Harris}, {Galvez}, {Benz}, {Fitzsimmons},
  {Green}, {Michel}, {Valsecchi}, {Paetzold}, {Haeusler}, \&
  {Carnelli}}]{donquijote-2006}
{Harris}, A.~W. {et~al.} 2006, in COSPAR Meeting, Vol.~36, 36th COSPAR
  Scientific Assembly, 2002

\bibitem[{Holsapple \& Housen(2012)}]{holsapple-housen-2012}
Holsapple, K.~A., \& Housen, K.~R. 2012, \textit{Icarus}

\bibitem[{Jutzi \& Michel(2014)}]{jutzi-michel-2014}
Jutzi, M., \& Michel, P. 2014, \textit{Icarus}, 229, 247

\bibitem[{Mazanek {et~al.}(2014)Mazanek, Merrill, Belbin, Reeves, Earle, Naasz,
  \& Abell}]{arm-2014}
Mazanek, D.~D., Merrill, R.~G., Belbin, S.~P., Reeves, D.~M., Earle, K.~D.,
  Naasz, B.~J., \& Abell, P.~A. 2014, in AIAA SPACE 2014 Conference and
  Exposition, San Diego, CA

\bibitem[{Scheeres {et~al.}(2015)Scheeres, McMahon, Jones, \&
  Doostan}]{scheeres-2015}
Scheeres, D.~J., McMahon, J.~W., Jones, B.~A., \& Doostan, A. 2015, in
  Aerospace Conference, 2015 IEEE, IEEE, 1--7

\bibitem[{{Spoto} {et~al.}(2014){Spoto}, {Milani}, {Farnocchia}, {Chesley},
  {Micheli}, {Valsecchi}, {Perna}, \& {Hainaut}}]{spoto-et-al-2014}
{Spoto}, F., {Milani}, A., {Farnocchia}, D., {Chesley}, S.~R., {Micheli}, M.,
  {Valsecchi}, G.~B., {Perna}, D., \& {Hainaut}, O. 2014,
  \textit{A}\&\textit{A}, 572, A100, 1408.0736

\bibitem[{{Sugimoto} {et~al.}(2014){Sugimoto}, {Radice}, {Ceriotti}, \&
  {Sanchez}}]{sugimoto-et-al-2014}
{Sugimoto}, Y., {Radice}, G., {Ceriotti}, M., \& {Sanchez}, J.~P. 2014,
  \textit{AcA}, 103, 333

\bibitem[{Yoshikawa {et~al.}(2006)Yoshikawa, Fujiwara, \&
  Kawaguchi}]{hayabusa-2006}
Yoshikawa, M., Fujiwara, A., \& Kawaguchi, J. 2006, Proceedings of the
  International Astronomical Union, 2, 323

\end{thebibliography}
%

\end{document}